# Interactive Data Exploration with Smart Drill-Down (Extended Version)

Manas Joglekar, Hector Garcia-Molina, *Member, IEEE,* Aditya Parameswaran, *Member, IEEE,*

**Abstract**—We present *smart drill-down*, an operator for interactively exploring a relational table to discover and summarize "interesting" groups of tuples. Each group of tuples is described by a *rule*. For instance, the rule $(a, b, \star, 1000)$ tells us that there are a thousand tuples with value $a$ in the first column and $b$ in the second column (and any value in the third column). Smart drill-down presents an analyst with a list of rules that together describe interesting aspects of the table. The analyst can tailor the definition of interesting, and can interactively apply smart drill-down on an existing rule to explore that part of the table. We demonstrate that the underlying optimization problems are NP-HARD, and describe an algorithm for finding the approximately optimal list of rules to display when the user uses a smart drill-down, and a dynamic sampling scheme for efficiently interacting with large tables. Finally, we perform experiments on real datasets on our experimental prototype to demonstrate the usefulness of smart drill-down and study the performance of our algorithms.

## 1 INTRODUCTION

Analysts often use OLAP (Online Analytical Processing) operations such as drill down (and roll up) [7] to explore relational databases. These operations are very useful for analytics and data exploration and have stood the test of time; all commercial OLAP systems in existence support these operations. (Recent reports estimate the size of the OLAP market to be $10+ Billion [21].)

However, there are cases where drill down is ineffective; for example, when the number of distinct values in a column is large, vanilla drill down could easily overwhelm analysts by presenting them with too many results (i.e., aggregates). Further, drill down only allows us to instantiate values one column at a time, instead of allowing simultaneous drill downs on multiple columns—this simultaneous drill down on multiple columns could once again suffer from the problem of having too many results, stemming from many distinct combinations of column values.

In this paper, we present a new interaction operator that is an extension to the traditional drill down operator, aimed at providing *complementary* functionality to drill down in cases where drill down is ineffective. We call our operator *smart drill down*. At a high level, smart drill down lets analysts zoom into the more "interesting" parts of a table or a database, with fewer operations, and without having to examine as much data as traditional drill down. Note that our goal is *not* to replace traditional drill down functionality, which we believe is fundamental; instead, our goal is to provide auxiliary functionality which analysts are free to use whenever they find traditional drill downs ineffective.

In addition to presenting the new smart drill down operator, we present novel sampling techniques to compute the results for this operator *in an interactive fashion* on increasingly larger databases. Unlike the traditional OLAP setting, these computations require no pre-materialization, and can be implemented within or on top of any relational database system.

We now explain smart drill-down via a simple example.

**Example 1.** *Consider a table with columns 'Department Store', 'Product', 'Region' and 'Sales'. Suppose an analyst queries for tuples where Sales were higher than some threshold, in order to find the best selling products. If the resulting table has many tuples, the analyst can use traditional drill down to explore it. For instance, the system may initially tell the analyst there are 6000 tuples in the answer, represented by the tuple* ($\star$, $\star$, $\star$, 6000, 0)*, as shown in Table 1. The $\star$ character is a wildcard that matches any value in the database. The Count attribute can be replaced by a Sum aggregate over some measure column, e.g., the total sales. The right-most Weight attribute is the number of non-$\star$ attributes; its significance will be discussed shortly. If the analyst drills down on the Store attribute (first $\star$), then the operator displays all tuples of the form* ($X$, $\star$, $\star$, $C$, 1)*, where $X$ is a Store in the answer table, and $C$ is the number of tuples for $X$ (or aggregate sales for $X$).*

*Instead, when the analyst uses smart drill down on Table 1, she obtains Table 2. The* ($\star$, $\star$, $\star$, 6000) *tuple is expanded into* 3 *tuples that display noteworthy or interesting drill downs. The number* 3 *is a user specified parameter, which we call $k$.*

*For example, the tuple* (Target, bicycles, $\star$, 200, 2) *says that there are* 200 *tuples (out of the 6000) with Target as the first column value and bicycle as the second. This fact tells the analyst that Target is selling a lot of bicycles. The next tuple tells the analyst that comforters are selling well in the MA-3 region, across multiple stores. The last tuple states that Walmart is doing well in general over multiple products and regions. We call each tuple in Table 2 a* rule *to distinguish it from the tuples in the original table that is being explored. Each rule summarizes the set of tuples that are described by it. Again, instead of Count, the operator can display a Sum aggregate, such as the total Sales.*

*Suppose after seeing the results of Table 2, the analyst wishes to dig deeper into the Walmart tuples represented by the last rule.*

- *M. Joglekar is with Stanford University. E-mail: joglekarmanas@gmail.com*
- *H. Garcia-Molina is with Stanford University. E-mail: hector@cs.stanford.edu*
- *A. Parameswaran is with the University of Illinois (UIUC). E-mail: adityagp@illinois.edu*

| Store | Product | Region | Count | Weight |
|---|---|---|---|---|
| $\star$ | $\star$ | $\star$ | 6000 | 0 |

*TABLE 1: Initial summary*

| Store | Product | Region | Count | Weight |
|---|---|---|---|---|
| $\star$ | $\star$ | $\star$ | 6000 | 0 |
| $\triangleright$ Target | bicycles | $\star$ | 200 | 2 |
| $\triangleright$ $\star$ | comforters | MA-3 | 600 | 2 |
| $\triangleright$ Walmart | $\star$ | $\star$ | 1000 | 1 |

*TABLE 2: Result after first smart drill down*

*The analyst may want to know which states Walmart has more sales in, or which products they sell the most. In this case, the analyst clicks on the Walmart rule, obtaining the expanded summary in Table 3. The three new rules in this table provide additional information about the* 1000 *Walmart tuples. In particular, one of the new rules shows that Walmart sells a lot of cookies; the others show it sells a lot of products in the regions CA-1 and WA-5.*

*When the analyst clicks on a rule $r$, smart drill down expands $r$ into $k$ sub-rules that as a set are deemed to be "interesting." There are three factors that make a rule set interesting. One is if it contains rules with high Count, since the larger the count, the more tuples are summarized. A second factor is if the rules have high weight (number of non-$\star$ attributes). For instance, the rule (Walmart, cookies, AK-1,* 200, 3*) is more interesting than (Walmart, cookies, $*$,* 200, 2*) since the former tells us the high sales are concentrated in a single region. The third factor is diversity: For example, if our set already has the rule (Walmart, $\star$, $\star$,* 1000, 1*), we would rather have add rule (Target, bicycles, $\star$,* 200, 2*) than (Walmart, bicycles, $\star$,* 200, 2*) since the former rule describes tuples that are not described by the first rule.*

In this paper we describe how to combine or blend these three factors in order to obtain a single desirability score for a set of rules. Our score function can actually be tuned by the analyst (by specifying how weights are computed), providing significant flexibility in what is considered a good set of rules. We also present an efficient optimization procedure to maximize score, invoked by smart drill down to select the set of $k$ rules to display.

**Relationship to Other Work.** Compared to traditional drill down, our smart drill down has two important advantages:

- Smart drill down limits the information displayed to the most interesting $k$ facts (rules). With traditional drill down, a column is expanded and *all* attribute values are displayed in arbitrary order. In our example, if we drill down on say the store attribute, we would see all stores listed, which may be a very large number.
- Smart drill down explores several attributes to open up together, and automatically selects combinations that are interesting. For example, in Table 2, the rule (Target, bicycles, $\star$, 200, 2) is obtained after a single drill down; with a traditional approach, the analyst would first have to drill down on Store, examine the results, drill down on Product, look through all the displayed rules and then find the interesting rule (Target, bicycles, $\star$, 200, 2).

Note that in the example we only described one type of smart drill down, where the analyst selects a *rule* to drill down on (e.g., the Walmart rule going from Table 2 to Table 3). In Section 2.3 we describe another option where the analyst clicks on a $\star$ in a column to obtain rules that have non-$\star$ values in that column.

| Store | Product | Region | Count | Weight |
|---|---|---|---|---|
| $\star$ | $\star$ | $\star$ | 6000 | 0 |
| $\triangleright$ Target | bicycles | $\star$ | 200 | 2 |
| $\triangleright$ $\star$ | comforters | MA-3 | 600 | 2 |
| $\triangleright$ Walmart | $\star$ | $\star$ | 1000 | 1 |
| $\triangleright$ $\triangleright$ Walmart | cookies | $\star$ | 200 | 2 |
| $\triangleright$ $\triangleright$ Walmart | $\star$ | CA-1 | 150 | 2 |
| $\triangleright$ $\triangleright$ Walmart | $\star$ | WA-5 | 130 | 2 |

*TABLE 3: Result after second smart drill down*

Our work on smart drill down is related to table summarization and anomaly detection [29], [28], [30], [14]. These papers mostly focus on giving "surprising" information to the user, i.e., information that would minimize the Kullback-Liebler(KL) divergence between the resulting maximum entropy distribution and the actual value distribution. For instance, if a certain set of values occur together in an unexpectedly small number of tuples, that set of values may be displayed to the user. In contrast, our algorithm focuses on rules with high counts, covering as much of the table as possible. Thus our work can be thought of as complementary to anomaly detection. Furthermore, our summarization is couched in an interactive environment, where the analyst directs the drill down and can tailor the optimization criteria.

Our work is also related to pattern mining. Several pattern mining papers [36], [9], [39] focus on providing one shot summaries of data, and do not propose interactive mechanisms. Moreover, to the best of our knowledge, other pattern mining work is either not flexible enough [16], [34], [13], restricting the amount of tuning the user can perform, or so general [24] as to preclude efficient optimization. Our work also merges 'interesting pattern mining' into the OLAP framework. We discuss related work in detail in Section 7.

**Contributions.** Our chief contribution in this paper is the *smart drill down* operator, an extension of traditional drill down, aimed at allowing analysts to zoom into the more "interesting" parts of a dataset. In addition to this operator, we develop techniques to support this operator on increasingly larger datasets:

- *Basic Interaction:* We demonstrate that finding the optimal list of rules is NP-HARD, and we develop an algorithm to find the approximately optimal list of rules to display when the user performs a smart drill down operation.
- *Dynamic Sample Maintenance:* To improve response time on large tables, we formalize the problem of dynamically maintaining samples in memory to support smart drill down. We show that optimal identification of samples is once again NP-HARD, and we develop an approximate scheme for dynamically maintaining and using multiple samples of the table in memory.

We have developed a *fully functional and usable prototype tool* that supports the smart drill-down operator that was demonstrated at VLDB 2015 [20]. From this point on, when we provide result snippets, these will be screenshots from our prototype tool. Our prototype tool also supports traditional drill-down: smart drill-down can be viewed as a generalization of traditional drill-down (with the weighting function set appropriately). In Section 5.1, we compare smart drill-down with traditional drill-down and show that smart drill-down returns considerably better results.

Our tool and techniques are also part of a larger effort for building DATASPREAD [6], a data analytics system with a spreadsheet-based front-end, and a database-based back-end, combining the benefits of spreadsheets and databases.

**Overview of paper:**

- In Section 2, we formally define smart drill down. After that, we describe different schemes for weighting rules, and our interactive user interface.

- In Section 3, we present our algorithms for finding optimal sets of rules.
- In Section 4, we present our dynamic sampling schemes for dealing with large tables
- Based on our implemented smart drill down, in Section 5 we experimentally evaluate performance on real datasets, and show additional examples of smart drill down in action.
- Section 6 covers extensions of our work. We describe related work in Section 7, and conclude in Section 8.

## 2 FORMAL DESCRIPTION

We describe our formal problem in Section 2.1, describe different scoring functions in Section 2.2, and describe our operator interfaces in Section 2.3.

### 2.1 Preliminaries and Definitions

**Tables and Rules:** As in a traditional OLAP setting, we assume we are given a star or snowflake schema; for simplicity, we represent this schema using a single denormalized relational table, which we call $\mathcal{D}$. For the purpose of the rest of the discussion, we will operate on this table $\mathcal{D}$. We let $T$ denote the set of tuples in $\mathcal{D}$, and $C$ denote the set of columns in $\mathcal{D}$.

Our objective (formally defined later) is to enable smart drill downs on this table or on portions of it: the result of our drill downs are lists of *rules*. A *rule* is a tuple with a value for each column of the table. In addition, a rule has other attributes, such as count and weight (defined later) associated with it. The value in each column of the rule can either be one of the values in the corresponding column of the table, or $\star$, representing a wildcard character representing all values in the column. For a column with numerical values in the table, we allow the corresponding rule-value to be a range instead of a single value. The *trivial rule* is one that has a $\star$ value in all columns. The *Size* of a rule is defined as the number of non-starred values in that rule.

**Coverage:** A rule $r$ is said to *cover* a tuple $t$ from the table if all non-$\star$ values for all columns of the rule match the corresponding values in the tuple. We abuse notation to write this as $t \in r$. At a high level, we are interested in identifying rules that cover many tuples. We next define the concept of subsumption that allow us to relate the coverage of different rules to each other.

We say that rule $r_1$ is a *sub-rule* rule $r_2$ if and only if $r_1$ has no more stars than $r_2$ and their values match wherever they both have non-starred values. For example, rule ($a$, $\star$) is a sub-rule of ($a$, $b$). If $r_1$ is a sub-rule of $r_2$, then we also say that $r_2$ is a *super-rule* of $r_1$. If $r_1$ is a sub-rule of $r_2$, then for all tuples $t$, $t \in r_2 \Rightarrow t \in r_1$.

**Rule Lists:** A *rule-list* is an ordered list of rules returned by our system in response to a smart drill down operation. When a user drills down on a rule $r$ to know more about the part of the table covered by $r$, we display a new rule-list below $r$. For instance, the second, third and fourth rule from Table 2 form a rule-list, which is displayed when the user clicks on the first (trivial) rule. Similarly, the second, third and fourth rules in Table 3 form a rule-list, as do the fifth, sixth and seventh rules.

**Scoring:** We now define some additional properties of rules; these properties help us score individual rules in a rule-list.

There are two portions that constitute our scores for a rule as part of a rule list. The first portion dictates how much the rule $r$ "covers" the tuples in $\mathcal{D}$; the second portion dictates how "good" the rule $r$ is (independent of how many tuples it covers). The reason why we separate the scoring into these two portions is that they allow us to separate the inherent goodness of a rule from how much it captures the data in $\mathcal{D}$.

We now describe the first portion: we define *Count*($r$) as the total number of tuples $t \in T$ that are covered by $r$. Further, we define *MCount*($r$, $R$) (which stands for 'Marginal Count') as the number of tuples covered by $r$ but not by any rule before $r$ in the rule-list $R$. A high value of $MCount$ indicates that the rule not only covers a lot of tuples, but also covers parts of the table not covered by previous rules. We want to pick rules with a high value of $MCount$ to display to the user as part of the smart drill down result, to increase the coverage of the rule-list.

Now, onto the second portion: we let $W$ denote a function that assigns a non-negative *weight* to a rule based on how good the rule is, with higher weights assigned to better rules. The weighting function does not depend on the specific tuples in $\mathcal{D}$, but could depend on the number of $\star$s in $r$, the schema of $\mathcal{D}$, as well as the number of distinct values in each column of $\mathcal{D}$. A weighting function is said to be *monotonic* if for all rules $r_1$, $r_2$ such that $r_1$ is a sub-rule of $r_2$, we have $W(r_1) \leq W(r_2)$; we focus on monotonic weighting functions because we prefer rules that are more "specific" rather than those that are more "general" (thereby conveying less information). We further describe our weighting functions in Section 2.2.

Thus, the total score for our list of rules is given by

$$\text{Score}(R) = \sum_{r \in R} \underbrace{MCount(r, R)}_{\text{coverage of } r \text{ in } \mathcal{D}} \times \underbrace{W(r)}_{\text{weight of } r}$$

Our goal is to choose the rule-list of a given length that maximizes total score.

We use $MCount$ rather than Count in the above equation to ensure that we do not redundantly cover the same tuples multiple times using multiple rules, and thereby increase coverage of the table. If we had defined total score as $\sum_{r \in R} \text{Count}(r)W(r)$, then our optimal rule-list could contain rules that repeatedly refer to the most 'summarizable' part of the table. For instance, if $a$ and $b$ were the most common values in columns $A$ and $B$, then for some weighting functions $W$, the summary may potentially consist of rules $(a, b, \star)$, $(a, \star, \star)$, and $(\star, b, \star)$, which tells us nothing about the part of the table with values other than $a$ and $b$.

Our smart drill downs still display the Count of each rule rather than the $MCount$. This is because while $MCount$ is useful in the rule selection process, Count is easier for a user to interpret. In any case, it would be a simple extension to display MCount in another column.

**Formal Problem:** We now formally define our problem:

**Problem 1.** *Given a table $T$, a monotonic weighting function $W$, and a number $k$, find the list $R$ of $k$ rules that maximizes*

$$\sum_{r \in R} W(r) \times MCount(r, R)$$

*for one of the following smart drill down operations:*

- [*Rule drill down*] *If the user clicked on a rule $r'$, then all $r \in R$ must be super-rules of $r'$*
- [*Star drill down*] *If the user clicked on a $\star$ on column $c$ of rule $r'$, then all $r \in R$ must be super-rules of $r'$ and have a non-$\star$ value in column $c$*

Throughout this paper, we use the *Count* aggregate of a rule to display to the user. We can also use a *Sum* of values over a given 'measure column' instead. We discuss how to modify our algorithms to use $Sum$ instead of $Count$ in Section 6.

### 2.2 Weighting Rules

We now describe our weighting function $W$ that is used to score individual rules. At a high level, we want our rules to be as descriptive of the table as possible, i.e. given the rules, it should be as easy as possible to reproduce the table. We consider a general family of weighting functions, that assigns for each rule $r$, a weight $W(r)$ depending on how expressive the rule is (i.e., how much information it conveys). We mention some canonical forms for function $W(r)$; later, we specify the full family of weighting functions our techniques can handle:

**Size Weighting Function:** $W(r) = |\{c \in C \mid r(c) \neq \star\}|$ : Here we set weight equal to the number of non-starred values in the rule $r$ i.e. the *size* of the rule. For example, in Table 2, the rule (Target, bicycles, $\star$) has weight 2.

To get an intuitive feel for this scoring function, imagine we are trying to reconstruct the table from the rules. Since we have rule ($a$, $b_1$) with $MCount$ 100, we are going to get a hundred of the table's tuples from this rule. For those hundred tuples, out of the 200 total values to be filled (2 per tuple, since there are 2 columns), all 200 values will already have been filled (since the rule specifies both columns). Thus, this rule contributes 200 to the score. For the rule ($a$, $\star$), there are 900 table tuples, and the $a$ value will be pre-filled for those tuples. Thus, 900 slots of these tuples have been pre-filled, and so the rule contributes 900 to the total. Thus, this scoring function can be thought of as the number of values that have been pre-filled in the table by our rule-list. Since having more of the table pre-filled is better, maximizing the score gives us a desirable set of rules.

**Bits Weighting Function:** $W(r) = \sum_{c \in C : r(c) \neq \star} \lceil \log_2(|c|) \rceil$ where $|c|$ refers to the number of distinct possible values in column $c$. This function weighs each column based on its inherent complexity, instead of equally like the Size function.

**Other Weighting Functions:** Even though we have given two example weighting functions here, our algorithms allow the user to leverage any weighting function $W$, subject to two conditions:

- Non-negativity: For all rules $r$, $W(r) \geq 0$.
- Monotonicity: If $r_1 \geq r_2$, then $W(r_1) \leq W(r_2)$. Monotonicity means that a rule that is less descriptive than another must be assigned a lower weight.

A weight function can be used in several ways, including expressing a higher preference for a column (by assigning higher weight to rules having a non-$\star$ value in that column), or expressing indifference towards a column (by adding zero weight for having non-$\star$ value in that column).

### 2.3 Smart drill down Operations

When the user starts using a system equipped with the smart drill down operator, they first see a table with a single trivial rule as shown in Table 1. At any point, the user can click on either a rule, or a star within a rule, to perform a 'smart drill down' on the rule. Clicking on a rule $r$ causes $r$ to expand into the highest-scoring rule-list consisting of super-rules of $r$. By default, the rule $r$ expands into a list of 3 rules, but this number can be changed by the user. The rules obtained from the expansion are listed directly below $r$, ordered in decreasing order by weight (the reasoning behind the ordering is explained in Section 3).

Instead of clicking on a rule, the user can click on a $\star$, say in column $c$ of rule $r$. This will also cause rule $r$ to expand into a rule-list, but this time the new displayed rules are guaranteed to have non-$\star$ values for in column $c$. Finally, when the user clicks on a rule that has already been expanded, it reverses the expansion operation, i.e. collapses it. For example, clicking on the walmart rule in Table 3 would take the user back to Table 2. This operation is equivalent to a traditional roll up, but for smart drill downs instead of traditional drill downs.

## 3 Smart drill down Algorithms

We now describe online algorithms for implementing the smart drill down operator. We assume that all columns are categorical (so numerical columns have been bucketized beforehand). We further discuss bucketization of numerical attributes in Section 6.

### 3.1 Problem Reduction and Important Property

When the user drills down on a rule $r'$, we want to find the highest scoring list of rules to expand rule $r'$ into. If the user had clicked on a $\star$ in a column $c$, then we have the additional restriction that all resulting rules must have a non-$\star$ value in column $c$. We can reduce Problem 1 to the following simpler problem by removing the user-interaction based constraints:

**Problem 2.** *Given a table $T$, a monotonic weight function $W$, and a number $k$, to find the list $R$ of $k$ rules that maximizes the total score given by :*

$$Score(R) = \sum_{r \in R} W(r) MCount(r, R)$$

Problem 1 with parameters $(T, W, k)$ can be reduced to Problem 2 as follows:

1) [Rule drill down] If the user clicked on rule $r$ in Problem 1, then we can conceptually make one pass through the table $T$ to filter for tuples covered by rule $r$, and store them in a temporary table $T_r$. Then, we solve Problem 2 for parameters $(T_r, W, k)$.
2) [Star drill down] If the user clicked on a $\star$ in column $c$ of rule $r$, then we first filter table $T$ to get a smaller table $T_r$ consisting of tuples from $T$ that are covered by $r$. In addition, we change the weight function $W$ from Problem 1 to a weight function $W'$ such that : For any rule $r'$, $W'(r') = 0$ if $r'$ has a $\star$ in column $c$, and $W'(r') = W(r')$ otherwise. Then, we solve Problem 2 for parameters $(T_r, W', k)$.

As a first step towards solving Problem 2, we show that the rules in the optimal list must effectively be ordered in decreasing order by weight. Note that the weight of a rule is independent of its $MCount$. The $MCount$ of a rule is the number of tuples that have been 'assigned' to it, and each tuple assigned to rule $r$ contributes $W(r)$ to the total score. Thus, if the rules are not in decreasing order by weight in a rule list $R$, then switching the order of rules in $R$ transfers some tuples from a lower weight rule to a higher weight rule, which can increase total score.

**Lemma 1.** *Let $R$ be a rule-list. Let $R'$ be the rule-list having the same rules as $R$, but ordered in descending order by weight. Then $Score(R') \geq Score(R)$.*

*Proof.* The score of rule list $R$ is given by

$$\text{Score}(R) = \sum_{r \in R} W(r) \times MCount(r, R)$$

For each tuple $t$, let $TOP(t, R)$ denote the first rule in rule-list $R$ that covers $t$. Then the MCount of rule $r$ in $R$ is simply $\sum_{t \in T: TOP(t,R)=r} 1$. Thus Score can be rewritten as:

$$\text{Score}(R) = \sum_{t \in T} W(TOP(t, R))$$

where we set $W(TOP(t)) = 0$ when $t$ is not covered by any rule in $R$. Now say two rule lists $R$, $R'$ have the same rules, but $R'$ has rules in decreasing order by weight. For any tuple $t$ covered by $R$, let $r'$ be the highest weight rule in $R$ that covers $t$. Let $r$ be the first rule in $R$ that covers $t$. Then $TOP(t, R) = r$, $TOP(t, R') = r'$ and $W(r') \geq W(r)$, so $W(TOP(t, R')) \geq W(TOP(t, R))$. Adding these inequalities for all $t$ covered by $R$ gives us $\text{Score}(R') \geq \text{Score}(R)$ as required. □

Thus, it is sufficient to restrict our attention to rule-lists that have rules sorted in decreasing order by weight. Or equivalently, we can define Score for a *set* of rules as follows:

**Definition 2.** *Let $R$ be a set of rules. Then the Score of $R$ is $Score(R) = Score(R')$ where $R'$ is the list of rules obtained by ordering the rules in the set $R$ in decreasing order by weight.*

This gives us a reduced version of Problem 2:

**Problem 3.** *Given a table $T$, a monotonic weight function $W$, and a number $k$, find the set (not list) $R$ of $k$ rules which maximizes Score($R$) as defined in Definition 2.*

The reduction from Problem 2 to Problem 3 is clear. We now first show that Problem 3, and consequently Problem 1 and Problem 2 are NP-HARD, and then present an approximation algorithm for solving Problem 3.

### 3.2 NP-Hardness for Problem 3

We reduce the well known NP-HARD *Maximum Coverage Problem* (MCP) to a special case of Problem 3; thus demonstrating the NP-HARDness of Problem 3. MCP is given below:

**Problem 4.** *Given a universe set $U$, an integer $k$, and a set $S = \{S_1, S_2, ...S_m\}$ of subsets of $U$ (so each $S_i \subset U$), find $S' \subset S$ such that $|S'| = k$, which maximizes Coverage($S'$) $= |\bigcup_{s \in S'} s|$.*

Thus, the goal of MCP is to find a set of $k$ of the given subsets of $U$ whose union 'covers' as much of $U$ as possible. We can reduce an instance of MCP (with parameters $U, k, S$) to an instance of Problem 3, which gives us the following lemma:

**Lemma 2.** *Problem 3 is* NP-HARD.

*Proof.* Consider a table with $|U|$ rows (one per element of $U$) and $m$ columns (one per $S_j$, that has value 1 in row $i$, column $j$ if the $i^{th}$ belongs to $S_j$ and 0 otherwise. And consider a weighting function with $W(r) = 1$ if there is at least one 1 in $r$, and 0 otherwise. Let $k \leq m$.

Then a rule with multiple 1s is clearly dominated (or matched) by a subrule with only one 1. Moreover, a rule-list that has two rules with a 1 in the same column is dominated by one that that 1s in different columns. Then the score of a rule list that has rules with 1s in column set $C \subset \{1, 2, \ldots m\}$ with $|C| = k$ has score equal to the size of union $\bigcup_{c \in C} S_c$. Thus maximizing score is equivalent to maximizing the size of the union of $k$ sets, which is the Maximum Coverage Problem. Thus the NP-Hardness of Maximum Coverage implies that Score maximization is NP-hard. □

### 3.3 Algorithm Overview

Given that Problem 3 is NP-HARD, we now present our algorithms for approximating the solution to it. The problem consists of finding a set of rules, given size $k$, that maximizes Score.

The next few sections fully develop the details of our solution:

- We show that the Score function is *submodular*, and hence an approximately optimal set can be obtained using a greedy algorithm. At a high level, this greedy algorithm is simple to state. The algorithm runs for $k$ steps; we start with an empty rule set $R$, and then at each step, we add the next best rule that maximizes Score
- In order to find the rule $r$ to add in each step, we need to measure the impact on Score for each $r$. This is done in several passes over the table, using ideas from the a-priori algorithm [4] for frequent item-set mining.

In some cases, the dataset may still be too large for us to return a good rule set in a reasonable time; in such cases, we may want to run our algorithm on a sample of the table rather than the entire table. In Section 4, we describe a scheme for maintaining multiple samples in memory and using them to improve response time for different drill down operations performed by the user. Our sampling scheme dynamically adapts to the current interaction scenario that the user is in; drawing from ideas in approximation algorithms and optimization theory.

### 3.4 Greedy Approximation Algorithm

**Submodularity:** We will now show that the Score function over sets of rules has a property called *submodularity*, giving us a greedy approximation algorithm for optimizing it.

**Definition 3.** *A function $f : 2^S \to \mathbb{R}$ for any set $S$ is said to be submodular if and only if, for every $s \in S$, and $A \subset B \subset S$ with $s \notin A$: $f(A \cup \{s\}) - f(A) \geq f(B \cup \{s\}) - f(B)$*

Intuitively, this means that the marginal value of adding an element to a set $S$ cannot increase if we add it to a superset of $S$ instead. For monotonic non-negative submodular functions, it is well known that the solution to the problem of finding the set of a given size with maximum value for the function can be found approximately in a greedy fashion.

**Lemma 3.** *For a given table $T$, the Score function over sets $S$ of rules, defined by the following is submodular:*

$$Score(S) = \sum_{r \in S} MCount(r, S) W(r)$$

*Proof.* Let $R(S)$ be the rule list obtained by sorting $S$ in descending weight order. For each tuple $t$, let $TOP(t, S)$ denote the first rule in rule-list $R(S)$ that covers $t$. By definition of $R(S)$, $r$ must also have the highest weight out of all rules in $S$ that cover $t$. Then $MCount(r, S)$ is simply $\sum_{t \in T: TOP(t,S)=r} 1$. Thus Score can be rewritten as:

$$\text{Score}(S) = \sum_{t \in T} W(TOP(t, S))$$

Now let $S \subsetneq S'$ and $s \notin S$. Then to prove submodularity, we simply need to prove that

$$\text{Score}(S \cup \{s\}) - \text{Score}(S) \geq \text{Score}(S' \cup \{s\}) - \text{Score}(S')$$

. This gives us $\text{Score}(S \cup \{s\}) - \text{Score}(S) = \sum_{t \in T} W(TOP(t, S \cup \{s\})) - W(TOP(t, S))$ and an analogous equation for $S'$. Consider two cases:

1) $W(TOP(t, S' \cup \{s\})) - W(TOP(t, S')) > 0$: This means $s$ covers $t$ and has higher weight than any rule in $S'$ that covers $t$. That is, $TOP(t, S' \cup \{s\}) = s$. Since $S \subseteq S'$, we must have $W(TOP(t, S')) \geq W(TOP(t, S))$, and also $TOP(t, S \cup \{s\}) = s$. Thus we have $W(TOP(t, S' \cup \{s\})) - W(TOP(t, S')) \leq W(TOP(t, S \cup \{s\})) - W(TOP(t, S))$.
2) $W(TOP(t, S' \cup \{s\})) - W(TOP(t, S')) = 0$: In this case, clearly $W(TOP(t, S' \cup \{s\})) - W(TOP(t, S')) \leq W(TOP(t, S \cup \{s\})) - W(TOP(t, S))$ since the latter is non-negative.

Either way, we have $W(TOP(t, S' \cup \{s\})) - W(TOP(t, S')) \leq W(TOP(t, S \cup \{s\})) - W(TOP(t, S))$, and summing over all $t \in T$ gives us

$$\text{Score}(S \cup \{s\}) - \text{Score}(S) \geq \text{Score}(S' \cup \{s\}) - \text{Score}(S')$$

as required. □

**High-Level Procedure:** Based on the submodularity property, the greedy procedure, shown in Algorithm 1, has desirable approximation guarantees. Since Score is a submodular function of the set $S$, this greedy procedure is guaranteed to give us a score within a $1 - \frac{1}{e}$ factor of the optimum (actually, it is $1 - (\frac{k-1}{k})^k$ for $k$ rules, which is much better for small $k$).

The expensive step in the above procedure is where the Score is computed for every single rule. Given the number of rules can be as large as the table itself, this is very time-consuming.

Instead of using the procedure described above directly, we instead develop a "parameterized" version that will admit further approximation (depending on the parameter) in order to reduce computation further. We describe this algorithm next.

**Algorithm 1:** BRS

```
Input: k (Number of rules required), T (database table), m_w (max weight), W
       (weight function)
Output: S (Solution set of rules)
S = φ
for i from 1 to k do
    R_m = Find_best_marginal_rule(S, T, m_w, W)
    S = S ∪ {R_m}
return S
```

**Parametrized Algorithm:** Our algorithm pseudo-code is given in the box labeled Algorithm 1. We call our algorithm *BRS* (for **B**est **R**ule **S**et). BRS takes four parameters as input: the table $T$, the number of required rules $k$, a parameter $m_w$ (described in the next paragraph), and the weight function $W$.

The parameter $m_w$ stands for *Max Weight*. This parameter tells the algorithm to assume that all rules that appear in the optimal solution are going to have weight $\leq m_w$. Thus, if $S_o$ denotes set of rules with maximum score, then as long as $m_w \geq \max_{r \in S_o} W(r)$, BRS is guaranteed to return $S_o$. On the other hand if $m_w < W(r)$ for some $r \in S_o$, then there is a chance that the set returned by BRS does not contain $r$. BRS runs faster for smaller values of $m_w$, and may only return a suboptimal result if $m_w < \max_{r \in S_o} W(r)$. In practice, $\max_{r \in S_o} W(r)$ is usually small. This is because as the size (and weight) of a rule increases, its Count falls rapidly. The Count tends to decrease exponentially with rule size, while Weight increases linearly for common weight functions (such as Size). Thus, rules with high weight and size have very low count, and are unlikely to occur in the optimal solution set $S_o$. Our experiments in Section 5 also show that the weights of rules in the optimal set tend to be small. Later in the extensions section, we describe strategies for setting $m_w$ as well as other parameters.

BRS initializes the solution set $S$ to be empty, and then iterates for $k$ steps, adding the best marginal rule at each step. To find the best marginal rule, it calls a function to find the best marginal rule given the existing set of rules $S$.

### 3.5 Finding the Best Marginal Rule

In order to find the best marginal rule, we need to find the marginal values of several rules and then choose the best one. A brute-force way to do this would be to enumerate all possible rules, and to find the marginal value for each of those rules in a single pass over the data. But the number of possible rules may be almost as large as the size of the table itself, making this step very expensive in terms of computation and memory.

In order to avoid counting too many rules, we leverage a technique inspired by the *a-priori* algorithm for frequent itemset mining [4]. Recall that the a-priori algorithm is used to find all frequent itemsets that have a support greater than a threshold. Unlike the a-priori algorithm, our goal is to find the single best marginal rule. Since we only aim to find one rule at a time, our pruning power is significantly higher than a vanilla a-priori algorithm, and we terminate in much fewer passes over the dataset.

We compute the best marginal rule over multiple passes on the dataset, with the maximum number of passes equal to the maximum size of a rule. In the $j^{th}$ pass, we compute counts and marginal values for rules of size $j$. To give an example, suppose we had three columns $c_1$, $c_2$, and $c_3$. In the first pass, we would compute the counts and marginal values of all rules of size 1. In the second pass, instead of finding marginal values for all size 2 rules, we can use our knowledge of counts from the first pass to upper bound the potential counts and marginal values of size 2 rules, and be more selective about which rules to count in the second pass. For instance, suppose we know that the rule $(a, \star, \star)$ has a count of 1000, while $(\star, b, \star)$ has a count of 100. Then for any value $c$ in column $c_3$ we would know that the count of $(\star, b, c)$ is at most 100 because it cannot exceed that of $(\star, b, \star)$. This implies that the maximum marginal value of any super-rule of $(\star, b, c)$ having weight $\leq m_w$ is at most $100 m_w$. If the rule $(a, \star, \star)$ has a marginal value of 800, then the marginal value of any super-rule of $(\star, b, \star)$ cannot possibly exceed that of $(a, \star, \star)$. Since our aim is to only find the highest marginal value rule, we can skip counting for all super-rules of $(\star, b, \star)$ for future passes.

We now describe the function to find the best marginal rule. The pseudo-code for the function is in the box titled Algorithm 2. The function maintains a threshold $H$, which is the highest marginal value that has been found for any rule so far. The function makes several passes over the table (Step 3), counting marginal values for size $j$ rules in the $j^{th}$ pass. We maintain three sets of rules: $C$, the set of rules whose marginal values have been counted in all previous passes, $C_n$, the set of rules whose marginal values will be counted in the current pass, and $C_o$, the set of rules whose marginal values were counted in the previous pass. For the

**Algorithm 2:** Find best marginal rule

**Input:** $S$ (Current solution set), $T$ (database table), $m_w$ (max weight), $W$ (weight function)
**Output:** $R_m$ (Rule which adds the highest marginal value among rules with weight $\leq m_w$)
$H = 0$ ; /* Threshold for deciding if to count for a rule. */
$C = C_o = C_n = \phi$ ; /* Set of all, old and new candidate rules respectively. */
**for** *j from* 1 *to number of columns in* $T$ **do**
  **if** $j = 1$ **then**
    $C_n =$ all rules of size 1
  **else**
    $C_n =$ all size-$i$ super-rules of rules from $C_o$
  **foreach** $R \in C_n$ **do**
    $M = \infty$ ; /* Upper bound on marginal value count of $R$ */
    **foreach** *R-sub-rule* $R' \in C$ **do**
      $M = \min(M, \text{MarginalVal}(R') + \text{Count}(R')(m_w - W(R'))$
    **if** $(M < H)$ **then**
      $C_n = C_n \setminus \{R\}$ /* Delete $R$ if its max count is too small for $R$ to be in the solution */
  **if** $C_n = \phi$ **then**
    break;
  **foreach** $R \in C_n$ **do**
    Count$(R) = 0$ ; /* Initialize */
    MarginalValue$(R) = 0$ ; /* Initialize */
  **foreach** $t \in T$ **do**
    Let $R_S$ be the highest weight rule in $S$ that covers $t$
    **foreach** $R \in C_n$ *that covers* $t$ **do**
      Count$(R)$ ++
      MarginalValue$(R)$ + = $W(R) - \min(W(R), W(R_S))$
  $C = C \cup C_n$
  $C_o = C_n$
  $C_n = \phi$
  $H = \max_{R \in C}(\text{MarginalValue}(R))$
**return** $\text{argmax}_{r \in C}\text{MarginalValue}(r)$

first pass, we set $C_n$ to be all rules of size 1. Then we compute marginal values for those rules, and set $C = C_o = C_n$.

For the second pass onwards, we are more selective about which rules to consider for marginal value evaluation. We first set $C_n$ to be the set of rules of size $j$ which are super-rules of rules from $C_o$. Then for each rule $r$ from $C_n$, we consider the known marginal values of its sub-rules from $C$, and use them to upper-bound the marginal value of all super-rules of $r$, as shown in Step 3.3.2. Then we delete from $C_n$ the rules whose marginal value upper bound is less than the currently known best marginal value, since they have no chance of being returned as the best marginal rule. Then we make as actual pass through the table to compute the marginal value of the rules in $C_n$, as shown in Step 3.5. If in any round, the $C_n$ obtained after deleting rules is empty, then we terminate the algorithm and return the highest value rule.

The reader may be wondering why we did not simply count the score of each rule using a variant of the a-priori algorithm in one pass, and then pick the set of rules that maximizes score subsequently. This is because doing so will lead to a sub-optimal set of rules: by not accounting for the rules that have already been selected, we will not be able to ascertain the marginal benefit of adding an additional rule correctly.

**Runtime analysis:** Our algorithm calls the function for finding best marginal rule $k$ times. Thus the total computational cost is about $k$ times the cost of computing the best marginal rule. Let $c_n$ be the number of counts in $C_n$ at the end of the $n^{th}$ iteration (that is, after eliminating rules that have no chance of having the highest marginal value). Also, let $c_0 = 1$.

The cost of the finding the best marginal rule can be split into three parts: (i) Enumerating all new rules to add to $C_n$ at the start of the $n^{th}$ iteration. (ii) Determining which of the rules in $C_n$ to eliminate (iii) Finding counts and marginal values for the remaining $c_n$ rules in $C_n$. The number of super-rules we can consider for (i) can be no larger than $c_{n-1}|T|$, since each rule in $C_{n-1}$ can give us at most $|T|$ new candidate super-rules for $C_n$. The cost of (ii) is $O(|C_n|)$. Finally, the cost of (iii) is again $c_n|T|$, since we have two nested for loops over $T$ and $C_n$ respectively. Thus, our total cost is $O(|T| \sum_i c_i)$.

In the worst case, we may need to evaluate all possible rules; since the total possible number of rules is $|T||C|$ where $C$ is the set of columns, our cost is at most $O(k|C||T|^2))$. However, note that this worst case is unlikely to occur when we use the Size or Bits weighting function, because the count of a rule decreases rapidly with its size, making it unlikely that we will need to find counts for its super-rules.

If fact, say that values in columns occur independently, and let $x$ be the frequency of the most common value (note $x < 1$). Then the number of candidate rules is bounded by $c_i \leq c_1 x^i$, where $c_1$ is the number of distinct values in the table. This gives us $\sum_i c_i \leq c_1 \sum_i x^i \leq c_1/(1-x)$. Thus our cost is now only $O(k|T|c_1/(1-x))$ for some constant $x$. If $c_1$, $k$ and $x$ are taken to be constants (i.e. they do not scale with table size), then our cost is $O(|T|)$.

**Approximation ratio:** If $m_w$ is higher than the actual weight of every rule in the best rule list, then the only approximation ratio we incur is that in the greedy approximation, which is a multiplicative factor of up to $1 - 1/e$. On the other hand, if our $m_w$ estimate is smaller than the actual maximum weight $m_w^*$ of a rule in the best list, then we may end up picking a sub-optimal rule in the best marginal rule function.

We now bound the sub-optimality of our chosen rule, for the case where $m_w < m_w^*$. Suppose the best rule to pick is $r^*$, and we pick rule $r$. Since we estimated $r^*$'s weight to be at most $m_w$, we would have estimated $r^*$ to have marginal value $\text{Est}(r^*) = \sum_{t \in r^*}(m_w - W(t))$ where $t \in r^*$ iterates over all tuples covered by $r^*$, and $w(t)$ is the weight of the current highest weight rule covering $t$. Since we picked $r$ over $r^*$, it means MarginalValue$(r) \geq \text{Est}(r^*)$. Suppose $m$ is the highest weight of a rule selected in the previous steps of BRS (so $w(t) \leq m$ for all $t$). Then, the actual marginal value of $r^*$ is $\leq \frac{m_w - m}{m_w^* - m}\text{Est}(r^*)$. Thus, our pick of $r$ is at most a $\frac{m_w - m}{m_w^* - m}$ factor worse than the optimal pick. If we always choose our $m_w$ such that we have $\frac{m_w - m}{m_w^* - m} \geq a$, for some constant $a < 1$, then our final score is guaranteed to be within a $a(1 - 1/e)$ factor of the optimal score.

## 4 Dynamic Sampling for Large Tables

BRS makes multiple passes over the table in order to determine the best set of rules to display. This can be slow when the table is too large to fit in main memory. We can reduce the response time of smart drill down by running BRS on a sample of the table instead, trading off accuracy of our rules for performance. If we had obtained a sample $s$ by selecting each table tuple with probability $p$, and run BRS on $s$, then we multiply the count of each rule found by BRS, by $\frac{1}{p}$ to estimate its count over the full table.

In Section 4.1, we describe the problem of optimally allocating memory to different samples. We show that the problem is NP-Hard, and describe an approximate solution. In Section 4.3, we describe a component of the system called the *SampleHandler*, which is responsible for creating and maintaining multiple samples of different parts of the table in memory, and creating temporary samples for BRS to process.

### 4.1 Deciding what to sample

We are given a memory capacity $M$, and a minimum sample size $minSS$, both specified by the user. $minSS$ is the minimum number of tuples on which we are allowed to run BRS, without accessing the hard disk. A higher value of $minSS$ increases both accuracy and computation cost of our system.

At any point, we have a tree $U$ of rules displayed to the user. Initially, the tree consists of a single node corresponding to the empty rule. When the user drills down on a rule $r$, the sub-rules of $r$ obtained by running BRS are added as children of node $r$. Our system maintains multiple samples in memory, with one sample per rule in $U$. Specifically, for each rule $r \in U$, we choose an integer $n_r$, and create a sample $s_r$ consisting of $n_r$ uniformly randomly chosen tuples covered by $r$ from the table. Because of the memory constraint, we must have $\sum_{r \in U} n_r \leq M$.

When a user attempts to drill down on $r$, the SampleHandler takes all $n_r$ tuples from $s_r$, and also tuples covered by $r$ from samples $s_{r'}$ for all $r' \in U$ that are sub-rules of $r$. If the total number of such tuples is $\geq minSS$, then we run BRS on that set of tuples to perform the drill down. Note that this set of tuples forms a uniformly random sample of tuples covered by $r$. If not, then we need to access the hard disk to obtain more tuples covered by $r$.

Leaves of tree $U$ correspond to rules that may be drilled down on next. We assume there is a probability distribution over leaves, which assigns a probability that each leaf may be drilled down on next. This can be a uniform distribution, or a machine learned distribution using past user data. We aim to set sample sizes $n_r$ so as to maximize the probability that the next drill down can be performed without accessing disk.

If $r'$ is a sub-rule of $r$, and covers $x$ times as many rules as $r$, then it means that when drilling down on $r$, $s_{r'}$ can contribute around $\frac{n_{r'}}{x}$ tuples to the sample for $r$. We denote the ratio of selectivities $x$ by $S(r', r)$. $S(r', r)$ is defined to be 0 if $r'$ is not a sub-rule of $r$. If $r$ is to be drilled down on next, the total number of sample tuples we will have for $r$ from all existing samples is given by $ess(r) = \sum_{r' \in U} S(r', r) n_{r'}$. We can drill down on $r$ without accessing hard disk, if $ess(r) \geq minSS$. We now formally define our problem:

**Problem 5.** *Given a tree of rules $U$ with leaves $L$, a probability distribution $p$ over $L$, an integer $M$, and selectivity ratio $S(r_1, r_2)$ for each $r_1, r_2 \in U$, choose an integer $n_r \geq 0$ for each $r \in U$ so as to maximize :*

$$\sum_{r' \in L} p_{r'} I_{[ess(r') \geq minSS]}$$

*where the $I$'s are indicator variables, with : $\sum_{r \in U} n_r \leq M$*

Problem 5 is non-linear and non-convex because of the indicator variables. We can show that Problem 5 is NP-HARD using a reduction from the knapsack problem.

**Lemma 4.** *Problem 5 is* NP-HARD.

*Proof.* (Sketch; details in [20]) Suppose we are given an instance of the knapsack problem with $m$ objects, with the $i^{th}$ object having weight $w_i$ and value $v_i$. We are also given a weight limit $W$, and our objective is to choose a set of objects that maximizes value and has total weight $< W$. We will reduce this instance to an instance of Problem 5.

We first scale the $w_i$s and $W$ such that all $w_i$s are $< 1$. For Problem 5, we set $M$ to $(m+W) \times minSS$. Tree $U$ has $m$ special nodes $r_1, r_2, ...r_m$, and each $r_i$ has two leaf children $r_{i,1}, r_{i,2}$. All other leaves have expansion probability 0. The $S$ values are such that $\forall i \in \{1, 2, ..m\}, j \in \{1, 2\} : (S(x, r_{i,j}) \neq 0 \Rightarrow x = r_{i,j} || x = r_i)$. In reality, the $S$ values cannot be exactly zero, but can be made small enough for all practical purposes. Thus, $\forall 1 \leq i \leq m, j \in \{1, 2\} : ess(r_{i,j}) = n_{r_{i,j}} + n_{r_i} S(r_i, r_{i,j})$. In addition, $S(r_i, r_{i,1}) = 1$, and $S(r_i, r_{i,2}) = 1 - w_i$. Finally, $\forall i : p_{r_{i,1}} = \frac{2}{2m+1}$ and $p_{r_{i,2}} = \frac{v_i}{(2m+1)\sum_{j=1}^{m} v_i}$. So in any optimal solution, $\forall i : ess(r_{i,1}) = minSS$, and we've to decide which $i$'s also have $ess(r_{i,2}) = minSS$. Having $ess(r_{i,2}) = minSS$ requires consuming $w_i \times minSS$ extra memory and gives an extra $\frac{v_i}{(2m+1)(\sum_{j=1}^{m} v_j)}$ probability value. Thus, having $ess(r_{i,2}) = minSS$ is equivalent to picking object $i$ from the knapsack problem. Solving Problem 5 with the above $U$, $S$, $p$ and picking the set of $i$'s for which $ess(r_{i,2}) = minSS$ gives a solution to the instance of the knapsack problem. ☐

**Approximate DP Solution.** The problem as stated is NP-HARD, but with a simplifying assumption we can make the problem approximately solvable using Dynamic Programming. The assumption is: for each $r \in L$, we assume that its $ess$ can get tuples only from samples obtained for itself and its parent. That is, we set $S(r_1, r_2)$ to be zero if $r_1 \neq r_2$ and $r_2$ is not a child of $r_1$. This is similar to what we had for tree $U$ in our proof of Lemma 4. So now $ess(r') = n_{r'} + n_r S(r, r')$ where $r$ is the parent of $r'$.

Consider a rule $r_0 \in U \setminus L$. Let $M_{r_0}$ denote the set containing $r_0$ and all its leaf children. By our assumption, the number of tuples $n_{r'}$ for any rule$r' \in M_{r_0}$ only affects the $ess$ value of rules in $M_{r_0}$. This allows us to split the problem into multiple subproblems, with one subproblem per $M_{r_0}$. For each non-leaf rule $r_0$ and all its children, we compute all 'locally optimal' assignments of $n_r \mid r \in M_{r_0}$. Locally optimal means that we cannot get a higher 'probability value' $\sum_{r \in M_{r_0}} p_r I_{ess(r) \geq minSS}$ for the same 'sampling cost' $\sum_{r \in M_{r_0}} n_r$. Then we use dynamic programming to combine locally optimal solutions of different $M_{r_0}$s. We describe these steps in detail below:

Let $r_0 \in U \setminus L$. Let $r_1, r_2, ...r_d$ be the leaf children of $r_0$. For any child $r_i$, $n_{r_i}$ only contributes to its own $ess$, while $n_{r_0}$ contributes to the $ess$ of all children $r_1, ...r_d$. Given a value of $n_{r_0}$, in a locally optimal solution, each child $r_i$ must satisfy:

- If $n_{r_0} S(r_0, r_i) \geq minSS$, then $n_{r_i} = 0$ because otherwise, decreasing $n_{r_i}$ to 0 would lower its sampling cost without improving its probability score.
- If $n_{r_0} S(r_0, r_i) < minSS$, then either $n_{r_i} = 0$ or $n_{r_i} = minSS - n_{r_0} S(r_0, r_i)$. This is because if $n_{r_i}$ is between 0 and $minSS - n_{r_0} S(r_0, r_i)$, then we can decrease it to 0, and if it is $> minSS - n_{r_0} S(r_0, r_i)$, then we can decrease it to $minSS - n_{r_0} S(r_0, r_i)$. Both these decreases would decrease sampling cost without affecting probability score.

Thus, there are three kinds of children $r_i$: Those with (i) $ess \geq minSS$ but $n_{r_i} = 0$, (ii) $ess < minSS$ and $n_{r_i} = 0$, (iii) $ess = minSS$ and $n_{r_i} = minSS - n_{r_0} S(r_0, r_i)$. There are $3^d$ ways to assign each child to one of these categories, and each of those potentially gives us one locally optimal solution. Consider any such locally optimal solution $e$. For $e$ let children $r_{i_1}, r_{i_2}, ...r_{i_m}$ be in the first category, $r_{i_{m+1}}, ..r_{i_M}$ be in the second category, and $r_{i_{M+1}}, ..r_{i_d}$ in the third. Then the 'probability value' of solution $e$ is given by : $P(e) = \sum_{j=1}^{i_M} p_j$, and its 'Sampling Cost' is

$$S(e) = \frac{minSS}{S(r_0, r_{i_m})} + \sum_{j=i_m+1}^{i_M} minSS - \frac{minSS}{S(r_0, r_{i_j})}$$

Thus, there are $\leq 3^d$ locally optimal solutions; $d$ is usually small ($\leq k$), even when $U$ is big. So we enumerate all locally optimal solutions and find their sampling cost and probability scores.

The

$A\left[i+1\right]\left[j\right] = \max(A\left[i\right]\left[j\right], \max_{e \in E_{i+1}}(A\left[i\right]\left[j - S(e)\right] + P(e)))$

Dynamic programming solves this in $O(DS3^d)$ time.

### 4.2 Alternative Convex-Optimization based solution

We noted earlier that Problem 5 is NP-Hard, but can be approximately solved with an additional simplifying assumption regarding the $S(r_1, r_2)$ values. Instead of making this simplification, we can make the problem convex (and hence tractable) with two different simplifications. The first simplification is, we modify our objective function to use hinge-loss instead of a step function. That is, our new objective function to maximise is

$$\sum_{r' \in L} p_{r'} \min\left(1, \frac{ess(r')}{minSS}\right)$$

Here we assume that it is acceptable to run our algorithm on samples smaller than $minSS$, though we still prefer bigger sample sizes upto $minSS$. The other simplification we make is assuming that sample sizes are real numbers instead of integers. After determining optimal sample sizes, we can round them up to get integer sample sizes. This will increase the memory usage by at most $|U|$, the number of nodes in displayed tree, which is negligible compared to the memory capacity $M$, or $minSS$.

In addition, in order to express our problem as a convex minimization problem, we negate the objective function and aim to minimize it (which is equivalent to maximizing the original objective function). Thus, our new optimization problem becomes

**Problem 6.** *Given a tree of rules $U$ with leaves $L$, a probability distribution $p$ over $L$, an integer $M$, and selectivity ratio $S(r_1, r_2)$ for each $r_1, r_2 \in U$, choose a real number $n_r \geq 0$ for each $r \in U$ so as to minimize :*

$$\sum_{r' \in L} p_{r'} max\left(-1, -\frac{ess(r')}{minSS}\right)$$

*subject to :*

$$\sum_{r \in U} n_r \leq M$$

The constraint is linear in the $n_r$ variables, and hence convex. Each $ess$ value is a linear function of the $n_r$s, which makes $-\frac{ess(r')}{minSS}$ convex. The constant function $-1$ is convex as well. Since the maximum of two convex functions is convex, Problem 6 is a convex minimization problem, which means that its local optimum is also its global optimum. Thus, we can initialize all $n_r$s to 0 and then use stochastic gradient descent (or any other local optimization technique) to find their optimum values.

The main weakness of this approach is that the hinge-loss objective rewards values of $ess < minSS$, which may lead us to all leaves having large $ess$ values that are nonetheless less than $minSS$, and thus gives lower quality count estimates than required by the user.

**Additional optimizations:** There are some additional minor optimizations we can make to reduce the memory cost per sample, allowing us to store more and bigger samples. Suppose we have a sample $s$, and say its filter rule $f_s$ has value $v$ in column $c$. Then we know that each tuple $t$ in $T_s$ must also have value $v$ in column $c$, since it is covered by $f_s$. So we do not need to explicitly store the column $c$ value of any tuple in $T_s$. We only need to store the tuple values of columns that have a $\star$ value in $f_s$. In addition, we may have a tuple occur in multiple samples. Instead of storing the entire tuple repeatedly, we could create a dictionary of common tuples, and only store a pointer to the tuple's dictionary entry in $T_s$.

**Setting $minSS$:** Suppose a rule $r$ covers $x$ fraction of the tuples of $T$ i.e. $x|T|$ tuples. Say we have a uniform random sample $s$ of $T$. The samples has size $|T_s|$, and let $X_{r,s}$ be the random variable denoting the number of tuples of $T_s$ covered by $r$. Then $E\left[X_{r,s}\right] = x|T_s|$, and $\text{Dev}(X_{r,s}) \approx \sqrt{|T_s|x(1-x)}$. In order to get a good estimate of $x$ (and hence of $\text{Count}(r) = x|T|$), we want $E\left[X_{r,s}\right] >> \text{Dev}(X_{r,s})$. That is, $x|T_s| >> \sqrt{|T_s|x(1-x)} \Leftrightarrow \frac{x|T_s|}{1-x} >> 1$.

We want to set the parameter minSS such that we get good count estimates for rules when using a sample of size $|T_s| = minSS$. If a rule displayed in our summary has covers $x$ fraction of the tuples, we want minSS to be at least $\rho\frac{1-x}{x}$, So the value of minSS must be at least $\rho\frac{1-x}{x}$ where $\rho$ is a constant chosen by us based on how accurate we want the count estimate to be. Moreover, since we want good Count estimates for all rules displayed in the summary, we want $minSS >> \rho\frac{1-x}{x}$ where $x$ is the minimum fraction of tuples covered by any of the rules displayed in our summary.

Thus, a reasonable value of minSS can be found by bounding $\frac{1-x}{x}$. This is hard to do for arbitrary weighting functions, but we can do it for the Size weighting function. Let $c$ be the column with the fewest distinct values. Say it has $|c|$ values. Then the rule that has the most frequent value of $c$, and $\star$ everywhere else, must have a score of at least $\frac{|T|}{|c|}$. For example, if the table has $10000$ tuples in all, and there is a 'Education' column that has 5 possible values, then the most frequent value of Education must occur at least 2000 times. So the rule with the most frequent value for Education, and $\star$s elsewhere, must have a score of at least 2000.

The highest scoring rule can have weight at most $|C|$ (the total number of columns). Since the score of the highest scoring rule is at least $\frac{|T|}{|c|}$, its Count must be at least $\frac{|T|}{|C||c|}$. Thus if minSS is significantly larger than $|C||c|$, then the Count of the first few highest scoring rules should be well-approximated in a sample of size more than minSS. For example, if $|T| = 10000$, $|c| = 5$, $|C| = 10$, then we want $minSS >> 5 \times 10$.

### 4.3 Design of the SampleHandler

We now describe the design of the *SampleHandler*, which given a certain memory capacity $M$, and a minimum sample size $minSS$, creates, maintains, retrieves, and removes samples, in response to user interactions on the table. It uses algorithms from Section 4.1 to decide which samples to create, as we will see below.

At all points, the SampleHandler maintains a set of samples in memory. For instance, it may keep a sample of tuples used to expand the first (trivial) rule, and another sample used to expand the rule last clicked on by the user. Each sample $s$ is represented as a triple: (a) A 'filter' rule $f_s$, (b) a scaling factor $N_s$ and (c) a set $T_s$ of tuples from the table. The set $T_s$ consists of a $\frac{1}{N_s}$ uniformly sampled fraction of tuples covered by $f_s$. The scaling factor $N_s$ is used to translate the count of a rule on the sample into an estimate of the count over the entire table. The sum of $|T_s|$ over all samples $s$ is not allowed to exceed capacity $M$ at any point.

Whenever the user drills down on a rule $r$, our system calls the SampleHandler with argument $r$, which returns a sample $s$ whose filter value is given by $f_s = r$ and has $|T_s| \geq minSS$. Thus, the $T_s$ of the returned sample consists of a uniformly random set of tuples covered by $r$. The SampleHandler also computes $N_s$ when a sample is created. Then we run BRS on sample $s$ (with a modified weight function in case the user clicked on a $\star$) to obtain the list of rules to display. The counts of the rules on the sample are multiplied by $N_s$ before being displayed, to get estimated counts on the entire table. In addition, since the sample is uniformly random, we can also compute confidence intervals on the estimated count of each displayed rule, although we do not currently display the confidence intervals.

When the SamplerHandler gets called with argument $r$, it needs to find or create a sample with $r$ as the filter rule. At the beginning when it gets called with the empty rule as an argument, there are no samples in memory and it must make a pass through the data to generate a sample. Creating a new sample by making a pass through the table is called **Create** (further described below). At later stages, when there are potentially multiple samples available, there are multiple mechanisms it could use to return a sample for rule $r$:

- **Find:** If the SampleHandler finds an existing sample $s$ in memory, which has $r$ as its filter rule (i.e. $f_s = r$) and at least $minSS$ tuples ($|T_s| \geq minSS$, then it simply returns sample $s$. BRS can then be run on $s$.
- **Combine:** If **Find** doesn't work i.e., if the SampleHandler cannot find an existing sample with filter $r$ and $\geq minSS$ tuples, then it looks at all existing samples $s'$ such that $f_{s'}$ is a sub-rule of $r$. If the set of all tuples that are covered by $r$, from all such $T_{s'}$'s combined, exceeds $minSS$ in size, then we can simply treat that set as our sample for $r$. Tuples that are covered by $r$, from the combination of $T_{s'}$s, follow a uniform distribution. That is, each table tuple $t$ that is covered by $r$ is equally likely to appear in a $T_{s'}$.
- **Create:** If **Combine** doesn't work either, then the SampleHandler needs to create a new sample $s$ with $f_s = r$ by making a pass through the table. Making a pass can be expensive for big tables, so we only use **Create** when **Find** and **Combine** cannot be used. We can use reservoir sampling [26], [35] to get a uniformly random sample of given size in a single pass through the table.

  When the SamplerHandler uses **Create** for a rule $r$, it needs to access the hard disk to make a pass through the entire table. Since accessing the hard disk and making a pass through the entire table is usually a bottleneck, it can also do things like creating samples for rules other than $r$, and augmenting existing samples, in the same pass. Hence, we assume that in a **Create** phase, the SampleHandler not only creates one new sample for $r$, but also uses the algorithm from Section 4.1 to determine the new optimal allocation of memory $n_r$ for each displayed rule $r$. Then in a single pass, it creates a sample of size $n_r$ for each displayed $r$.

**Pre-fetching:** When the user clicks on rule $r$ (or on a $\star$ in $r$), we need to get a sample, run the BRS, and display a rule-list to the user. If we use **Find** or **Combine**, then we can display the rule-list much faster because we don't have to read the entire table. But after expanding $r$, there is a high chance that the user goes further and drills down on one of the sub-rules $r'$ of $r$. We may not be able to use **Find** or **Combine** on $r'$ with the existing samples. So while the user is busy reading the current rule-list obtained from drilling down on $r$, we can start running the algorithm from Section 4.1 in the background, and then making a pass through the table to create a new samples. That way, when the user expands the next rule $r'$, there will be a high chance of a sample being pre-fetched for $r'$, increasing the chance that we can use **Find** or **Combine** on $r'$ and reducing our response time. In addition, while we are making the pass in the background, we can find the exact counts for currently displayed rules (which only have estimated counts shown), and update them when our pass is complete.

## 5 EXPERIMENTS

We have implemented a fully-functional interactive tool instrumented with the smart drill down operator, having a web interface. We now describe our experiments on this tool with real datasets.

**Datasets.** The first dataset, denoted 'Marketing', contains demographic information about potential customers [1]. A total of $N = 9409$ questionnaires containing 502 questions were filled out by shopping mall customers in the San Francisco Bay area. This dataset is the summarized result of this survey. Each tuple in the table describes a single person. There are 14 columns, each of which is a demographic attribute, such as annual income, gender, marital status, age, education, and so on. Continuous values, such as income, have been bucketized in the dataset, and each column has up to 10 distinct values.

The columns (in order) are as follows: annual household income, gender, marital status, age, education, occupation, time lived in the Bay Area, dual incomes?, persons in household, persons in household under 18, householder status, type of home, ethnic classification, language most spoken in home.

The second dataset, denoted 'Census', is a US 1990 Census dataset from the UCI Machine Learning repository [5], consisting of about 2.5 million tuples, with each tuple corresponding to a person. It has 68 columns, including ancestry, age, and citizenship. Numerical columns, such as age, have been bucketized beforehand in the dataset. We use this dataset in Section 5.2 in order to study the accuracy and performance of sampling on a large dataset.

Unless otherwise specified, in all our experiments, we restrict the tables to the first 7 columns in order to make the result tables fit in the page. We use the current implementation of our the smart drill down operator, and insert cropped screenshots of its output in this paper. We set the $k$ (number of rules) parameter to 4, and $m_w$ to 5 for the Size weighting and 20 for the Bits weighting function (see Section 2.2). Memory capacity $M$ for the SampleHandler is set to 50000 tuples, and $minSS$ to 5000.

### 5.1 Qualitative Study

We first perform a qualitative study of smart drill down. We observe the effects of various user interface operations on the Marketing Dataset (the results are similar on the Census dataset), and then try out different weight functions to study their effects.

| | Gender | Marital Status | Age | Education | Occupation | Time in Bay Area | Count | Weight |
|---|---|---|---|---|---|---|---|---|
| | * | * | * | * | * | * | 8993 | 0 |
| > | Female | * | * | * | * | * | 4918 | 1 |
| > | Male | * | * | * | * | * | 4075 | 1 |
| > | Female | * | * | * | * | > 10 years | 2940 | 2 |
| > | Male | Never married | * | * | * | > 10 years | 980 | 3 |

*Fig. 1: Summary after clicking on the empty rule*

| | Gender | Marital Status | Age | Education | Occupation | Time in Bay Area | Count | Weight |
|---|---|---|---|---|---|---|---|---|
| | * | * | * | * | * | * | 8993 | 0 |
| > | Female | * | * | * | * | * | 4918 | 1 |
| >> | Female | * | * | High school | * | * | 1149 | 2 |
| >> | Female | * | * | Grades 9-11 | * | * | 605 | 2 |
| >> | Female | * | * | College graduate | * | * | 771 | 2 |
| >> | Female | * | * | 1-3 years college | * | * | 1712 | 2 |
| > | Male | * | * | * | * | * | 4075 | 1 |
| > | Female | * | * | * | * | > 10 years | 2940 | 2 |
| > | Male | Never married | * | * | * | > 10 years | 980 | 3 |

*Fig. 2: Star expansion on 'Education' Column*

#### 5.1.1 Testing the User Interface

We now present the rule-based summaries displayed as a result of a few different user actions. To begin with, the user sees an empty rule with the total number of tuples as the count. Suppose the user expands the rule. Then the user will see Figure 1. The first two new rules simply tell us that the table has $4918$ female and $4075$ male tuples. The next two rules also slightly more detailed, saying that there are $2940$ females who have been in the Bay Area for $> 10$ years, and $980$ males who have never been married and been in the Bay Area for $> 10$ years. Note that the latter two rules give very specific information which would require up to 3 user clicks to find using traditional drill down, whereas smart drill down displays that information to the user with a single click.

Now suppose the user decides to further explore the table, by looking at education related information of females in the dataset. Say the user clicks on the $\star$ in the 'Education' column of the second rule. This opens up Figure 2 that shows the number of females with different levels of education, for the $4$ most frequent levels of education among females. Instead of expanding the 'Education' column, if the user had simply expanded the third rule, it would have displayed Figure 3.

#### 5.1.2 Weighting functions

Our system can display optimal rule lists for any monotonic weighting function. By default, we assign a rule weight equal to its size. In this section, we consider other weighting functions.

We first try the 'Bits' weighting function, given by:

$$W(r) = \sum_{c \in C : r(c) \neq \star} \lceil \log_2(|c|) \rceil$$

where $|c|$ refers to the number of distinct values in column $c$. This function gives higher weight to rules that have non-$\star$ values in columns that have many distinct values. The rule summary for this weighting is in Figure 6 (contrast with Figure 1). Bits weighting gives low weight for non-$\star$ values in binary columns, like the gender column. Thus, this summary instead gives us information about the Marital Status/Time in Bay Area/Occupation columns instead of the Gender column like in Figure 1.

| | Gender | Marital Status | Age | Education | Occupation | Time in Bay Area | Count | Weight |
|---|---|---|---|---|---|---|---|---|
| | * | * | * | * | * | * | 8993 | 0 |
| > | Female | * | * | * | * | * | 4918 | 1 |
| > | Male | * | * | * | * | * | 4075 | 1 |
| >> | Male | Never married | * | * | * | * | 1897 | 2 |
| >> | Male | Married | * | * | * | * | 1368 | 2 |
| >> | Male | * | * | * | * | > 10 years | 2242 | 2 |
| > | Female | * | * | * | * | > 10 years | 2940 | 2 |

*Fig. 3: A rule expansion*

| | Gender | Marital Status | Age | Education | Occupation | Time in Bay Area | Count | Weight |
|---|---|---|---|---|---|---|---|---|
| | * | * | * | * | * | * | 8993 | 0 |
| > | * | * | 14-17 | * | * | * | 878 | 1 |
| > | * | * | 55-64 | * | * | * | 640 | 1 |
| > | * | * | 45-54 | * | * | * | 922 | 1 |
| > | * | * | 25-34 | * | * | * | 2249 | 1 |
| > | * | * | 35-44 | * | * | * | 1615 | 1 |
| > | * | * | 18-24 | * | * | * | 2129 | 1 |
| > | * | * | 64+ | * | * | * | 560 | 1 |

*Fig. 4: A regular drill down on Age*

The other weighting function we try is given by:

$$W(r) = \text{Min}(0, \text{Size}(r) - 1)$$

This gives us Figure 7. This weighting gives a $0$ weight to rules with a single non-$\star$ value, and thus forces the algorithm to finds good rules having at least $2$ non-$\star$ values. As a result, we can see that our system only displays rules having $2$ or $3$ non-$\star$ values, unlike Figure 1 which has two rules displaying the total number of males and females, that have size $1$.

A regular drill down is a special case of smart drill-down with the right weighting function and number of rules. Specifically, if we want to perform a regular drill down on a column $C$ that has $n$ distinct values, then we set the number of rules to be displayed ($k$) to $n$. The weighting function $W$ is set such that $W(r) = 1$ if rule $r$ has $C$ instantiated, and $0$ otherwise. This ensures that all displayed rules have $C$ instantiated (with no other column $C'$ instantiated unless there is a functional dependency from $C$ to $C'$), and that each displayed rule has a distinct value of $C$. This effectively gives us a regular drill-down on $C$. We use this to perform a drill down on the 'Age' column using our experimental prototype. The result is shown in Figure 4. We can contrast it with Figure 1; the latter gives information about multiple columns at once and only displays high count values. Regular drill down on the other hand, serves a complementary purpose by focusing on detailed evaluation of a single column.

### 5.2 Quantitative Study

The performance of our algorithm depends on various parameters, such as $m_w$ (the max weight) and $minSS$ (minimum required sample size). We now study the effects of these parameters on the computation time and accuracy of our algorithm. We use the Marketing and Census datasets. The Marketing dataset is relatively small with around $9000$ tuples, whereas the Census dataset is quite large, with $2.5$ million tuples. The accuracy of our algorithm depends on $m_w$ and $minSS$, rather than the underlying database size. The worst case running time for large datasets is close to the time taken for making one pass on the dataset. When we expand a rule using an existing sample in memory, the running time is small and only depends on $minSS$ rather than on the dataset size.

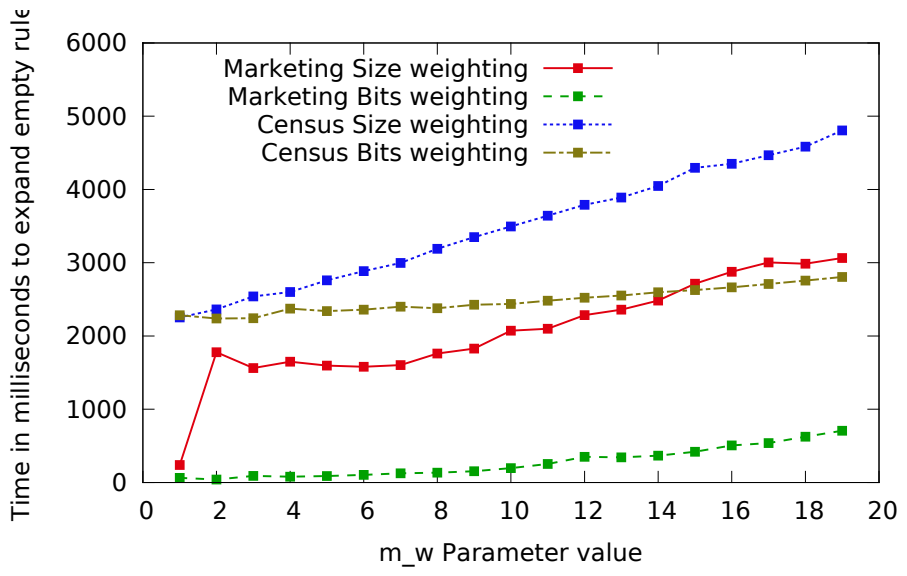

*Fig. 5: Running time for different values of parameter $m_w$*

| | Gender | Marital Status | Age | Education | Occupation | Time in Bay Area | Count | Weight |
|---|---|---|---|---|---|---|---|---|
| - | * | * | * | * | * | * | 8993 | 0 |
| + > | * | * | * | * | * | > 10 years | 5182 | 3 |
| + > | * | * | * | * | Professional / Managerial | * | 2820 | 4 |
| + > | * | Never married | * | * | Student | > 10 years | 742 | 10 |
| + > | * | Married | * | * | Professional / Managerial | > 10 years | 825 | 10 |

*Fig. 6: Bits scoring*

| | Gender | Marital Status | Age | Education | Occupation | Time in Bay Area | Count | Weight |
|---|---|---|---|---|---|---|---|---|
| - | * | * | * | * | * | * | 8993 | 0 |
| + > | Female | * | * | * | * | > 10 years | 2940 | 1 |
| + > | Male | Never married | * | * | * | > 10 years | 980 | 2 |
| + > | Female | Married | * | * | * | > 10 years | 1230 | 2 |
| + > | Male | Married | * | * | * | > 10 years | 823 | 2 |

*Fig. 7: Size minus one weighting*

### 5.2.1 Effects of $m_w$

Our algorithm for finding the best marginal rule takes an input parameter called $m_w$. The algorithm is guaranteed to find the best marginal rule as long as its weight is $\leq m_w$, but runs faster for smaller values of $m_w$. We now study the effect of varying $m_w$ on the speed of our algorithm running on a Dell XPS L702X laptop with 6GB RAM and an Intel i5 2.30GHz processor.

We fix a weighting function $W$, and a value of $m_w$. For that value of the $W$ and $m_w$ parameters, we find the time taken for expanding the empty rule. We repeat this procedure 10 times and take the average value of the running times across the 10 iterations. This time is plotted against $m_w$, for $W(r) = \text{Size}(r)$ and $W(r) = \sum_{c \in C: r(c) \neq \star} \lceil \log_2(|c|) \rceil$ in Figure 5. The figure shows that running time seems to be approximately linear in $m_w$.

For the Census dataset, the running time is dominated by time spent in making a pass through the 2.5 million tuples to create the first sample. The response time for the next user click should be quite small, as the sample created for the first expansion can usually be re-used for the next rule expansion.

The value of $m_w$ required to ensure a correct answer is equal to the maximum weight of a selected rule. Thus, for size scoring on the Marketing dataset, according to Figure 1, we require $m_w \geq 3$. For the second weighting function, according to Figure 6, the minimum required value of $m_w$ is 10. At these values of $m_w$, we see that the expansion takes 1.5 seconds and about 0.25 seconds respectively. Of course, the minimum value of $m_w$ we can use is not known to us beforehand. But even if we use more conservative values of $m_w$, say 6 and 20 respectively, the running times are about 1.5 and 0.5 seconds respectively.

### 5.2.2 Effects of $minSS$

We now study the effects of sampling parameter $minSS$. This parameter determines the minimum sample size on which we run BRS. Higher values of $minSS$ cause our system to use bigger samples, increasing the accuracy of count estimates for displayed rules, but also correspondingly increasing computation time.

We consider one value of $minSS$ and one weight function $W$ at a time. For those values of $minSS$ and $W$, we drill down on the empty rule and measure the time taken. We also measure the percent error in the estimated counts of the displayed rules. That is, for each displayed rule $r$, if the displayed (estimated) count if $c_1$ and the actual count (computed separately on the entire table) is $c_2$, then the percent error for rule $r$ is $\frac{100 \times |c_1 - c_2|}{c_2}$. We consider the average of percent errors over all displayed rules. For each value of $minSS$ and $W$, we drill down on the empty rule and find the computation time and percent error 50 times, and take the average value for time and error over those 50 iterations. This average time is plotted against $minSS$, for $W(r) = \text{Size}(r)$ and $W(r) = \sum_{c \in C: r(c) \neq \star} \lceil \log_2(|c|) \rceil$ in Figure 8(a). The average percent error is plotted against $minSS$, for $W(r) = \text{Size}(r)$ and $W(r) = \sum_{c \in C: r(c) \neq \star} \lceil \log_2(|c|) \rceil$ in Figure 8(b).

Figure 8(a) shows that sampling gives us noticeable time savings. The percent error decreases approximately as $\frac{1}{\sqrt{minSS}}$, which is again expected because the standard deviation of estimated $Count$ is approximately inversely proportional to the square root of sample size.

In addition, we measure the number of incorrect rules per iteration. If the correct set of rules to display is $r_1, r_2, r_3$ and the displayed set is $r_1, r_3, r_4$ then that means there is one incorrect rule. We find the number of incorrect displayed rules across 50 iterations, and display the average value in Figure 8(c). This number for the Marketing dataset is almost always 0 for the Size weighting function, and between 1 and 2 for the Bits weighting function. For the Census dataset, it is around 1 for $minSS \leq 1000$ and falls to about 0.3 for larger values of $minSS$. Note that even when we display an 'incorrect' rule, it is usually the $5^{th}$ or $6^{th}$ best rule instead of one of the top 4 rules, which still results in a reasonably good summary of the table.

### 5.2.3 Scaling properties of our algorithms

The computation time for a smart drill-down is linear in both the table size $|T|$ and in parameter $minSS$. That is, the runtime can be written as $a \times |T| + b \times minSS$ for some constants $a$ and $b$. In the worst-case where we cannot form a sample from main memory and need to re-create a sample, $a$ stands for the time taken to read data from hard disk. That is, $a \times |T|$ is the time taken to make a single scan over the table on disk. $b$ is bigger than $a$, because BRS makes multiple passes over the sample, while creating a sample only requires a single pass over the table.

When $|T|$ is small, the runtme is dominated by the $b \times minSS$ term, as seen for the Marketing Dataset in Figure 8(a). When $|T|$ is large relative to $minSS$, like for the Census Dataset, the runtime is dominated by $a \times |T|$ (this is when we need to create a fresh sample from hard disk). When we have a few million tuples, our total runtime is only a few seconds. But if the dataset

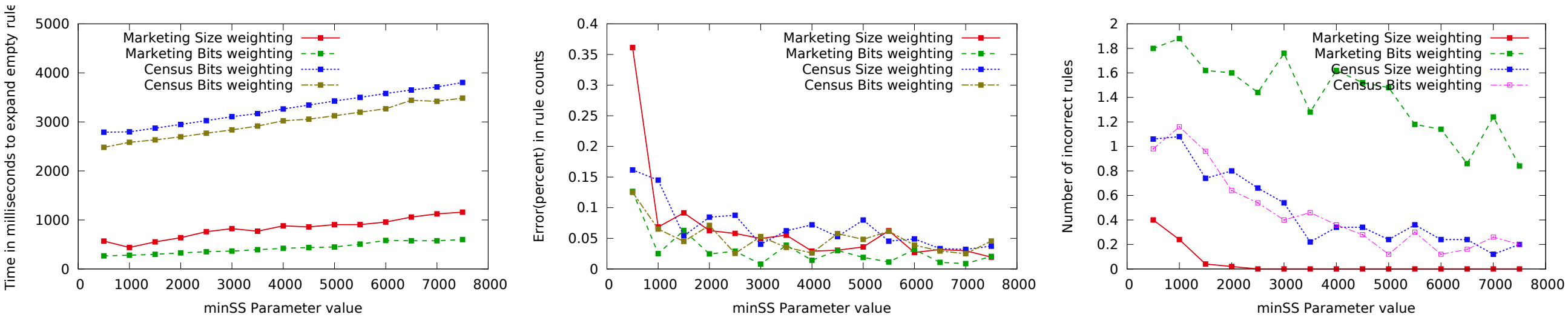

*Fig. 8: (a) Running time for different values of parameter $minSS$ (b) Error in Count for different values of parameter $minSS$ (c) Average number of incorrect rules for different values of parameter $minSS$*

contained billions of tuples, the process of reading the table to create a sample could itself take a very long time. To counteract this, we could preprocess the dataset by down-sampling it to only a million tuples, and perform the summarization on the million tuple sample (which also summarizes the billion tuple table).

## 6 Extensions

### 6.1 Setting parameters $W$, $k$, $m_w$

Our system allows the user to tune the smart drill-down by adjusting a number of parameters. Having a lot of tunable parameters can increase the difficulty of using a system by increasing the decision-making burden on the user. To counteract this, we now provide ways to guide the user while selecting appropriate parameter values.

Parameter $k$ is the number of new rules to display upon each smart drill-down. Large values of $k$ increase the run-time quadratically, and can also overwhelm the user with too much information. Very small values of $k$ may display too little information about the table. Fortunately, the BRS algorithm is incremental in nature. That is, in order to find the best rule list of size $k+1$, it first finds the best rule-list of size $k$, and then finds another rule to add to get a rule-list of size $k+1$. Thus instead of running the algorithm with a fixed value of $k$, it can start with an empty rule-list and keep adding rules to it, displaying new rules as they are found. This search for additional rules can stop when the user issues a new smart drill-down command to the system, or manually stops the search. Alternatively, we can set a time limit (of say 5 seconds) and display as many rules as we can find within that time limit.

$W$ is the weight function that determines which rules are interesting. This is a function specified by the user as a black box. Specifying an arbitrary function can be hard, so instead we hardcode some common Weight functions and allow the user to choose one from a drop-down menu. In addition, the user can express interest or disinterest in certain columns by telling the system to favor or ignore those columns, via the user interface. The system internally adjusts the weight function by increasing or decreasing the weight given to rules instantiating that column.

The $m_w$ parameter lets the user trade off the accuracy of the optimal rule-list and the running time. Ideally we want $m_w$ to equal the actual maximum weight of a rule in the optimal rule-list; this way we get full accuracy while also optimizing run-time. We cannot know the ideal value of $m_w$ in advance, but we can easily estimate it using sampling. We create a small random sample of tuples from the table, and run the BRS algorithm on it. Then the maximum weight $x$ of the output on the sample is likely to equal the maximum weight of the actual output. To account for sampling error, we can set $m_w$ to $2x$, which works well in practice.

**Generalizing our weight functions:** We now analyze a parametric family of weighting functions that generalizes our functions from Section 2.2 and provides intuition for them. Let $W(r) = \left(\sum_{c\in C} o_{r,c}w_c\right)^k$, where the $w_c$'s and $k$ are parameters of the weighting function, and $o_{r,c}$ ($o$ stands for occurrence) is 1 if $r(c) \neq \star$ and 0 otherwise. The Size and Bits weighting functions can be seen as special cases of the above function with $k=1$ and with $w_c$ equal to 1 for Size and $\log(|c|)$ for Bits.

We want to estimate the Score contribution of a rule according to this weight function, which requires estimating the count of the rule. Suppose that for each column $c$, the most frequent value occurs a $f_c$ fraction of the time. If the probability of a value occurring in a column of a tuple is independent of other values in the tuple, the $Count$ of rule $r$ is approximately $|T|\Pi_{c\in C} f_c^{o_{r,c}}$. Thus the score of a rule set containing $r$ alone can be approximated by:

$$S(r) = \left(\sum_{c\in C} o_{r,c}w_c\right)^k |T|\Pi_{c\in C} f_c^{o_{r,c}}$$

Now consider the rule that has the highest weight times count according to the above weighting function. We can approximate this rule by relaxing $o_{r,c}$ to a real number; that is by maximizing the estimated $S(r)$ subject to $0 \leq o_{r,c} \leq 1$. According to the KKT conditions [8], at the optimum $Weight \times Count$, the partial derivative of $W(r)$ with respect to $o_{r,c}$ must be $\leq 0$ for each $c$ where $o_{r,c} = 0$, equal to 0 if $0 < o_{r,c} < 1$ and $\geq 0$ otherwise. For each $c$, we have:

$$\frac{\partial S(r)}{\partial o_{r,c}} = S(r)\left(\frac{kw_c}{\sum_{c\in C} o_{r,c}w_c} + \ln f_c\right)$$

Thus, if $x = \frac{k}{\sum_{c\in C} o_{r,c}w_c}$, then the maximum scoring rule should instantiate columns where $\frac{\ln f_c}{w_c} > x$ and some of the ones with $\frac{\ln f_c}{w_c} = x$. Thus, for Size weighting, the highest score rule should select columns with high values of $\ln f_c$, i.e., the highest frequency values. In contrast, Bits weighting will select columns with highest values of $\log_{|c|}(f_c)$, i.e., values that are frequent relative to the number of distinct values in the column.

If we want the score to treat all columns equally, then the derivative must be 0 for all $c$. This implies that for all $c$, we must have $w_c \propto \ln f_c$. Notice that Bits satisfies the above criterion if we assume that values in each column are uniformly distributed (that is, $f_c = 1/|c|$). This provides an intuitive justification for why we might want to use the Bits weighting function.

We can also estimate the size of the optimal rule under the above weighting function. Specifically, the weighted fraction of columns with non-$\star$ values is given by $\frac{\sum_{c\in C} o_{r,c}w_c}{\sum_{c\in C} w_c}$ which equals $-k/(\sum_{c\in C} \ln f_c)$. Thus, if we wanted our highest score rule to have $s$ fraction of the columns get instantiated, we can achieve that by setting $k = -s \sum_{c \in C} \ln f_c$.

The above reasoning also lets us estimate the $m_w$ parameter we need to use. Specifically, the highest score rule above has weight $(\frac{-\ln f_c}{k w_c})^k$.

We can use the above estimate to guess the size of the max scoring rule for the Bits weighting function. But note that in practice, values are not distributed uniformly (so $f_c$ is actually higher than $1/|c|$) and are not independent of each other, due to correlations. As a result, our estimate of the maximum size is a significant under-estimate. In our experiments (Section 5.1), we observe that use the Bits weighting function actually gives us rules of size 3 on our datasets, while our estimate above is approximately $0.74$. As expected, the values displayed by the rules are ones that occur much more frequently relative to other values in that column.

### 6.2 Dealing with Numerical Attributes

Our framework assumes that all attributes are categorical. Attributes that have a large domain tend to have fewer tuples per value, and hence don't appear in rule summaries. Thus our algorithm does not summarize information about numerical attributes. But it can easily be modified to do so by bucketizing a numerical attribute and treating the bucket id as a categorical attribute. This is already done in our MD dataset, where numerical attributes like age are divided into buckets ($18 - 24$, $25 - 34$ and so on).

### 6.3 Using Sum instead of Count

We defined the total score of a rule-list using the marginal counts of rules in the list, and display the counts of rules in our table summary. However, if we have a numerical (i.e. a 'measure') column in the table, it is straightforward to extend our summary to the 'Sum' aggregate over that column. We can define $Sum$ and $MSum$ analogously to $Count$ and $MCount$, and $Score$ can be modified to use $MSum$ instead of $MCount$. We can then modify Algorithm 1 to maximize the new score by replacing $Count(r)$ by $Sum(r)$ and computing sum and marginal sum instead of count and marginal count in each pass over the table.

## 7 Related Work

There has been work on finding cubes for OLAP systems [29], [28], [30]. This and other work [25] focuses on finding values that occur more often or less often that expected from a max-entropy distribution. The work does not guarantee good coverage of the table, since it rates infrequent sets of values as highly as frequent ones. Some other data exploration work [31] focuses on finding attribute values that divide the database in equal sized parts, while we focus on values that occur as frequently as possible.

There is work on constructing 'explanation tables', sets of rules that co-occur with a given binary attribute of the table [14]. This work again focuses on displaying rules that will cause the resulting max entropy distribution to best approximate the actual distribution of values. A few vision papers [22], [12] suggest frameworks for building interactive data exploration systems. Some of these ideas, like maintaining user profiles, could be integrated into smart drill down. Reference [11] proposes an extension to OLAP drill-down that takes visualization real estate into account, by clustering attribute values. But it focuses on expanding a single column at a time, and relies on a given value hierarchy for clustering.

Some related work [18], [17] focuses on finding minimum sized Tableaux that provide improved support and confidence for conditional functional dependencies. There is some work [10], [23], [38], [15] on finding hyper-rectangle based covers for tables. In both these cases, the emphasis is on completely covering or summarizing the table, suffering from the same problems as traditional drill down in that the user may be presented with too many results. The techniques in the former case may end up picking rare "patterns" if they have high confidence, and in the latter case do not scale well to a $\geq 4$ attributes.

Several existing papers also deal with the problem of frequent itemset mining [4], [37], [19]. Vanilla frequent itemset mining is not directly applicable to our problem because the flexible user-specified objective function emphasizes coverage of the table rather than simply frequent itemsets. However, we do leverage ideas from the a-priori algorithm [4] as applicable. Several extensions have been proposed to the a-priori algorithm, including those for dealing with numerical attributes [33], [27]. We can potentially use these ideas to improve handing of numerical attributes in our work. Unlike our paper, there has been no work on dynamically maintaining samples for interaction in the frequent itemset literature, since frequent itemset mining is a one-shot problem.

There has also been plenty of work on pattern mining. Several papers [36], [9], [39] propose non-interactive schemes that attempt to find a one shot summary of the table. These schemes usually consume a large amount of time processing the whole table, rather than allowing the user to slowly steer into portions of interest. In contrast, our work is interactive, and includes a smart memory manager that can use limited memory effectively while preparing for future requests.

Our Smart Drill-Down operator is tunable because of the flexible weighting function, but the monotonicity of the weighting function and the use of $MCount$, still make it possible for us to get an approximate optimality guarantee for the rules we display. In contrast, much of the existing pattern mining work [16], [34], [13] is not not tunable enough, providing only a fixed set of interestingess parameters. On the other hand, reference [24] allows a fully general scoring function, necessitating the use of heuristics with no optimality guarantees, and very time consuming algorithms. A lot of pattern mining work [16], [39], [36] also focuses on itemsets rather than Relational Data, which does not allow the user to express interest in certain 'columns' over others.

We use sampling to find approximate estimates of rule counts. Various other database systems [2], [3] use samples to find approximate results to SQL aggregation queries. These systems create samples in advance and only update them when the database changes. In contrast, we keep updating our samples on the fly, as the user interacts with our system. There is work on using weighted sampling [32] to create samples favouring data that is of interest to a user, based on the user's history. In contrast, we create samples at run time in response to the user's commands.

## 8 Conclusion

We have presented a new data exploration operator called smart drill down. Like traditional drill down, it allows an analyst to quickly discover interesting value patterns (rules) that occur frequently (or that represent high values of some metric attribute) across diverse parts of a table.

We presented an algorithm for optimally selecting rules to display, as well as a scheme for performing such selections based on data samples. Working with samples makes smart drill down relatively insensitive to the size of the table.

Our experimental results on our experimental prototype show that smart drill down is fast enough to be interactive under various realistic scenarios. We also showed that the accuracy is high when sampling is used, and when the maximum weight ($m_w$) approximation is used. Moreover, we have a tunable parameter $minSS$ that the user can tweak to tradeoff performance of smart drill down for the accuracy of the rules.

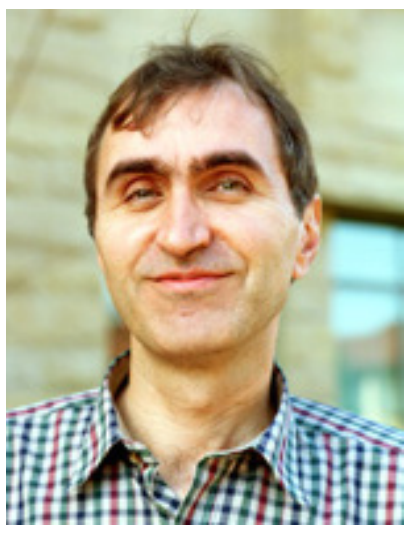

Hector Garcia-Molina is the Leonard Bosack and Sandra Lerner Professor in the Departments of Computer Science and Electrical Engineering at Stanford University, Stanford, California. He was the chairman of the Computer Science Department from January 2001 to December 2004. From 1997 to 2001 he was a member the President's Information Technology Advisory Committee (PITAC). From August 1994 to December 1997 he was the Director of the Computer Systems Laboratory at Stanford. From 1979 to 1991 he was on the faculty of the Computer Science Department at Princeton University, Princeton, New Jersey. His research interests include distributed computing systems, digital libraries and database systems. He received a BS in electrical engineering from the Instituto Tecnologico de Monterrey, Mexico, in 1974. From Stanford University, Stanford, California, he received in 1975 a MS in electrical engineering and a PhD in computer science in 1979. He holds an honorary PhD from ETH Zurich (2007). He is a Fellow of the Association for Computing Machinery and of the American Academy of Arts and Sciences; is a member of the National Academy of Engineering; received the 1999 ACM SIGMOD Innovations Award; is a Venture Advisor for Onset Ventures, is a member of the Board of Directors of Oracle, and is a member of the State Farm Technical Advisory Council.

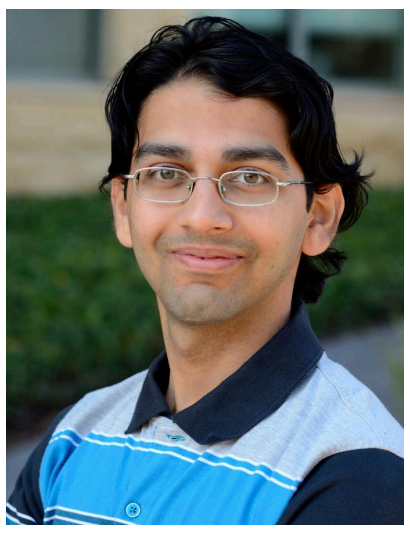

Manas Joglekar is a fifth year PhD student in the Computer Science Department at Stanford University. Manas' advisor is Prof. Hector Garcia-Molina, and his research interests include Crowdsourcing, Data Quality Evaluation and Database

Theory. Prior to that, he received a B. Tech in Computer Science and Engineering from the Indian Institute of Technology Bombay, in 2011. Manas is a recipient of International Mathematics Olympiad Silver (2007) and Bronze (2006) Medals, and two best-of-conference citations (ICDT 2016, ICDE 2016).

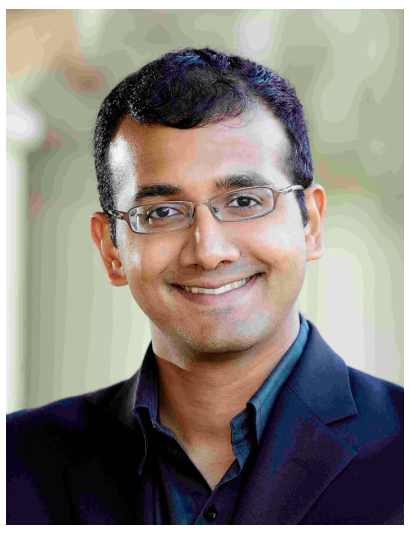

Aditya Parameswaran is an Assistant Professor in Computer Science at the University of Illinois (UIUC). He spent the 2013-14 year visiting MIT CSAIL and Microsoft Research New England, after completing his Ph.D. from Stanford University, advised by Prof. Hector Garcia-Molina. He is broadly interested in data analytics, with research results in human computation, visual analytics, information extraction and integration, and recommender systems. Aditya is a recipient of the Arthur Samuel award for the best dissertation in CS at Stanford (2014), the SIGMOD Jim Gray dissertation award (2014), the SIGKDD dissertation award runner up (2014), a Google Faculty Research Award (2015), the Key Scientific Challenges Award from Yahoo! Research (2010), four best-of-conference citations (VLDB 2010, KDD 2012, ICDE 2014, ICDE 2016), the Terry Groswith graduate fellowship at Stanford (2007), and the Gold Medal in Computer Science at IIT Bombay (2007). His research group is supported with funding from by the NIH, the NSF, and Google.